\documentclass[iop]{emulateapj}
\usepackage{verbatim}
\usepackage{gensymb}
\usepackage{graphicx}
\usepackage{array}
\usepackage{booktabs}
\usepackage{threeparttable}
\usepackage{natbib}
\newcommand{\otoprule}{\midrule[\heavyrulewidth]}
\newcolumntype{+}{>{\global\let\currentrowstyle\relax}}
\newcolumntype{^}{>{\currentrowstyle}}

\begin{document}

\title{The Distribution and Chemistry of H$_2$CO in the DM Tau Protoplanetary Disk}


\author{
Ryan A. Loomis\altaffilmark{1,2},
L. Ilsedore Cleeves\altaffilmark{3},
Karin I. {\"O}berg\altaffilmark{2},
Viviana V. Guzman\altaffilmark{2},
and
Sean M. Andrews\altaffilmark{4}}

\altaffiltext{1}{Corresponding author: rloomis@cfa.harvard.edu}
\altaffiltext{2}{Department of Astronomy, Harvard University, Cambridge, MA 02138}
\altaffiltext{3}{Department of Astronomy, University of Michigan, Ann Arbor, MI 48109}
\altaffiltext{4}{Harvard-Smithsonian Center for Astrophysics, Cambridge, MA 02138}


\begin{abstract}
H$_2$CO ice on dust grains is an important precursor of complex organic molecules (COMs). H$_2$CO gas can be readily observed in protoplanetary disks and may be used to trace COM chemistry.  However, its utility as a COM probe is currently limited by a lack of constraints on the relative contributions of two different formation pathways: on icy grain-surfaces and in the gas-phase.  We use archival ALMA observations of the resolved distribution of H$_2$CO emission in the disk around the young low-mass star DM Tau to assess the relative importance of these formation routes.  The observed H$_2$CO emission has a centrally peaked and radially broad brightness profile (extending out to 500 AU).  We compare these observations with disk chemistry models with and without grain-surface formation reactions, and find that both gas and grain-surface chemistry are necessary to explain the spatial distribution of the emission.  Gas-phase H$_2$CO production is responsible for the observed central peak, while grain-surface chemistry is required to reproduce the emission exterior to the CO snowline (where H$_2$CO mainly forms through the hydrogenation of CO ice before being non-thermally desorbed).  These observations demonstrate that both gas and grain-surface pathways contribute to the observed H$_2$CO in disks, and that their relative contributions depend strongly on distance from the host star.
\end{abstract}

\keywords{Protoplanetary Disks, Astrochemistry, Circumstellar Matter, ISM: Molecules, Radio Lines: ISM}

\section{Introduction}

The chemical structure of protoplanetary disks is set by a combination of gas and grain-surface chemistry, regulated by the temperature, density, and radiation properties of the disk \citep[e.g.][]{Aikawa_2002}. Radiation impinging upon the disk from both the central star and the interstellar radiation field results in strong radial and vertical temperature gradients, differentiating the disk into three chemical layers: a cold midplane, a hot atmosphere, and a warm molecular layer at intermediate heights, \citep{Aikawa_1996, Bergin_2007}. Gas-phase chemistry generally dominates above the midplane. Toward the cold midplane, and at larger radial distances from the star, the sequential freeze-out of volatiles regulates the chemical structure \citep{Qi_2013_Sci, Zhang_2013}.  Freeze-out of molecules depletes gas-phase abundances (affecting gas-phase chemistry), and determines where in the disk ice chemistry on grains can occur \citep[e.g.][]{Garrod_2007, Walsh_2010}.

Grain-surface chemistry is especially important for the production of complex organic molecules (COMs), and thus for the prebiotic potential of planetary bodies forming in the disk \citep[e.g.][]{Garrod_2008, Garrod_2013, Mumma_2011}. With one exception (CH$_3$CN)\citep{Oberg_2015}, COMs have not been possible to observe directly in disks. An alternative approach to estimating the COM abundance in disks is to constrain the chemistry that  produces them, by observing smaller organic molecules with a grain-surface formation pathway.

Of the molecules with predicted grain-surface formation routes, only H$_2$CO has been found to be readily observable in disks \citep{Dutrey_1997, Aikawa_2003, Thi_2004, Oberg_2010, Oberg_2011, Qi_2013_H2CO}. H$_2$CO is expected to form on grain-surfaces through the sequential hydrogenation of CO ice, a reaction which is well-studied in the laboratory \citep{Hiraoka_1994, Watanabe_2002, Fuchs_2009}. Atomic hydrogen for these reactions would likely come from ionization of H$_2$. Both the freeze-out of CO and the presence of H$_2$ ionization are well established in disks, based on observations of CO snowlines and of ions like HCO$^+$ \citep[e.g.][]{Qi_2011, Qi_2013_Sci, Mathews_2013, Kastner_1997}. A large uncertainty, however, is how efficiently H$_2$CO ice is desorbed into the gas-phase at different disk locations.

H$_2$CO can also form through gas-phase chemistry.\footnote{KIDA reaction network \citep{Wakelam_2015}} One important and well studied gas-phase reaction is between oxygen atoms and CH$_3$ \citep{Fockenberg_2002}, which is suspected to be a major contributor to H$_2$CO abundances in dense cores \citep{Guzman_2013}. 

A priori it is not clear whether gas or grain-surface reactions or both govern the H$_2$CO gas budget in disks, and thus the utility of H$_2$CO observations as a tracer of grain-surface organic chemistry is not well constrained. The spatial distribution of H$_2$CO should contain information on the contributions of these formation mechanisms. As the major gas-phase formation route of H$_2$CO has an activation barrier, it will contribute more heavily to the H$_2$CO abundance at higher densities (higher collision frequency) and temperatures, both of which are present in the inner disk. Gas-phase formation of H$_2$CO should therefore result in a centrally peaked abundance profile. In contrast, grain-surface formation of H$_2$CO requires CO to be frozen out on the grain-surfaces, and it should mainly contribute to H$_2$CO gas outside the CO snow line \citep{Henning_2008, Qi_2013_H2CO}.

Based on the qualitative differences in abundance profiles expected for different formation pathways (in the gas or on grain-surfaces), high spatial resolution imaging has the potential to constrain the chemical origin of H$_2$CO in disks. This approach was used by \citet{Qi_2013_H2CO} using Sub-Millimeter Array (SMA) observations of H$_2$CO toward HD 163296. Their observations of an H$_2$CO emission ring outside the CO snowline revealed the presence of grain-surface formation. In an earlier Plateau de Bure Interferometer (PdBI) study of the protoplanetary disk around the T Tauri star DM Tau, a resolved ring-shaped distribution of H$_2$CO was also interpreted as evidence of grain-surface chemistry exterior to the CO snowline \citep{Henning_2008}. 

In this letter, we revisit the large ($\sim$800 AU) and well-studied \citep[e.g.][]{Guilloteau_1994, Dutrey_2007, Oberg_2011, Teague_2015} disk around DM Tau with higher sensitivity Atacama Large (sub-)Millimeter Array (ALMA) observations to evaluate whether gas-phase chemistry significantly contributes to H$_2$CO abundances, or whether grain-surface chemistry truly dominates. We compare our observations with detailed chemical models of each formation scenario, constraining the importance of both gas-phase and grain-surface contributions.

\section{Observations}
\label{observe}

DM Tau was observed on 2012 October 23 in Band 7 as part of the ALMA cycle 0 project 2011.0.00629.S (PI: E. Chapillion).  The array configuration included 23 antennas with projected baseline lengths between 17.5 and 380 m (20-430 k$\lambda$), providing a synthesized beam of 0.65" x 0.45" (94 x 65 AU at 145 pc) at a PA of $\sim$-29$\degree$. The total integration time on-source was 35 minutes.  The correlator spectral setup consisted of four windows centered at 339.674, 340.366, 351.527, and 354.264 GHz, each with a total bandwidth of 468.75 MHz in 122.07 kHz (0.105 km s$^{-1}$) channels.  The observed 5$_{15}$ - 4$_{14}$ transition of H$_2$CO at 351.769 GHz is located 
in the third window.

The quasar J0510+180 was used for phase calibration, Callisto was used for flux calibration, and J0423-013 was used as the bandpass calibrator.  We self-calibrated the data in CASA version 4.2 using the DM Tau disk continuum emission.  

We imaged the continuum emission using the multi-frequency synthesis mode of the CASA \textit{CLEAN} task from a total bandwidth of 1.875 GHz, minus channels with significant line emission. For spectral line imaging, we continuum subtracted the visibilities in the uv plane and applied a restoring beam of 1.5" x 1.0" (217 x 145 AU at 145 pc) along the PA of the original beam to obtain a desirable balance of sensitivity and resolution. We generated an initial \textit{CLEAN} mask using a simple model of a disk in Keplerian rotation and then modified it to more closely fit the observed emission. The achieved rms is $\sim$0.16 mJy beam$^{-1}$ for the continuum and $\sim$10 mJy beam$^{-1}$ per H$_2$CO channel.

\section{Results\label{res}}

\subsection{H$_2$CO spectral line emission}

\begin{figure}[ht!]
\centering
\includegraphics[scale=0.87]{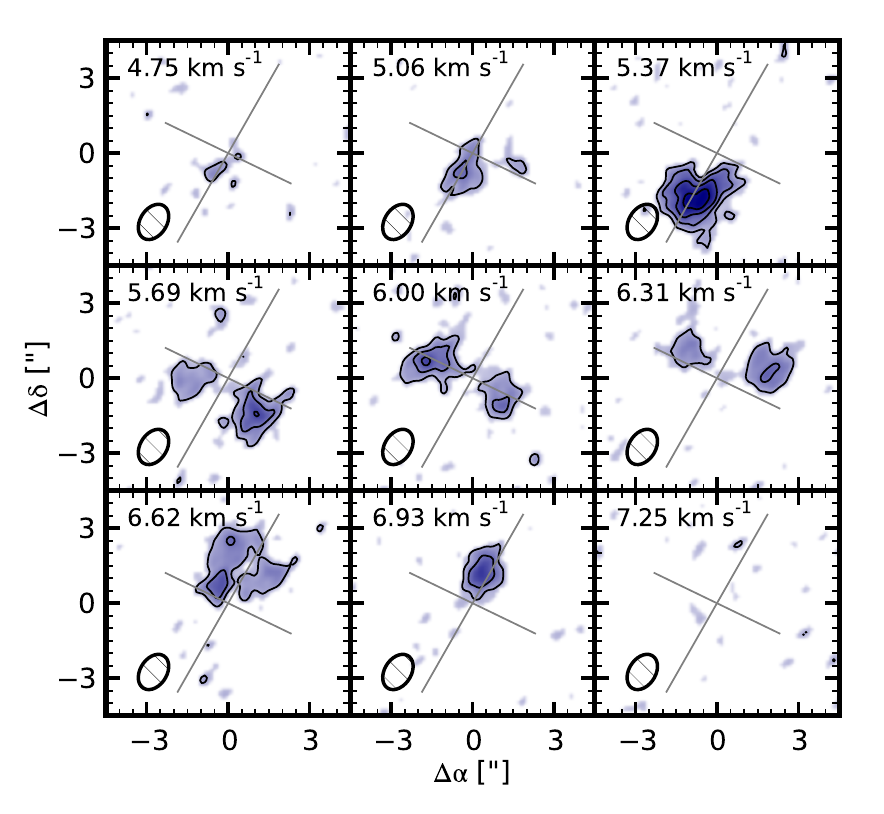}
\caption{{\small Channel map of H$_2$CO 5$_{15}$ - 4$_{14}$ emission toward DM Tau with contours of 8 (1$\sigma$) $\times$ [3,5,7] mJy beam$^{-1}$. The data were binned to a velocity resolution of 0.31 km s$^{-1}$ (i.e. 3 channels). The LSR channel velocity is shown in upper left and the synthesized beam in the lower left of each panel.\label{ch}}}
\end{figure}

\begin{figure*}[ht!]
\centering
\includegraphics{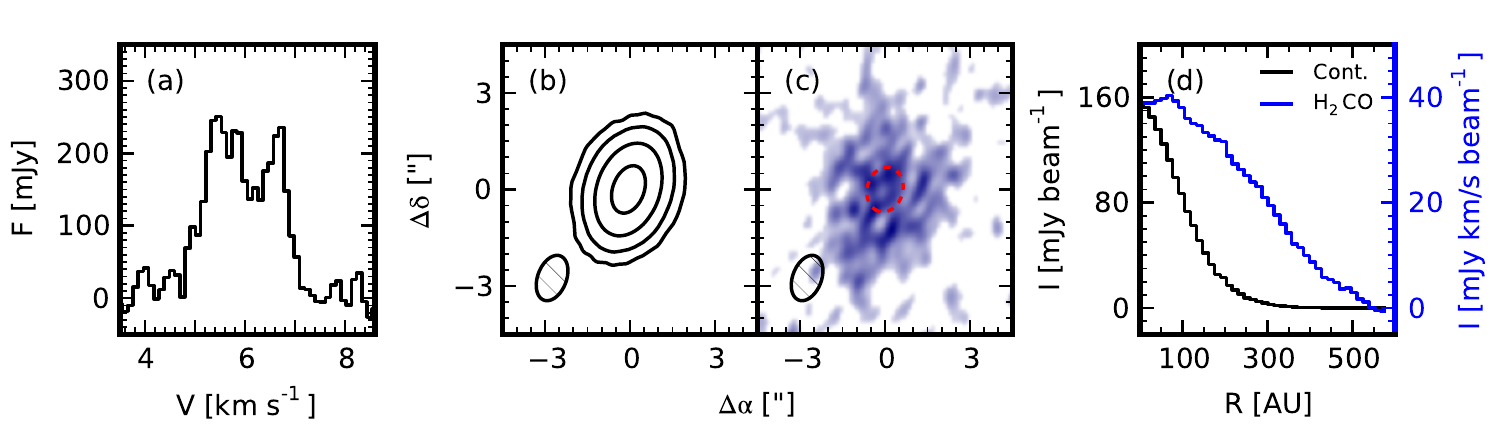}
\caption{{\small H$_2$CO spectrum, and images of observed dust and H$_2$CO emission from the disk around DM Tau. \textit{Panel (a):} Spectrum of H$_2$CO emission, integrated over the \textit{CLEAN} mask for each channel. \textit{Panel (b):} Continuum map, extracted from line-free channels as described in Section \ref{observe}. Contours denote 0.16 (1$\sigma$) $\times$ [10, 40, 160, 640] mJy beam$^{-1}$. \textit{Panel (c):} H$_2$CO emission, integrated from 4-8 km s$^{-1}$. The dashed red line denotes the expected CO snowline location. \textit{Panel (d):} Radial profiles of continuum and  H$_2$CO emission, deprojected and azimuthally averaged. The H$_2$CO profile is shown in blue with the scale to the right, while the continuum is shown in black with the scale to the left.}\label{mom}}
\end{figure*}

Figure \ref{ch} presents channel maps of the H$_2$CO observations.  The emission is consistent with Keplerian rotation about the continuum peak at $\alpha$J2000 = 04$^{h}$33$^{m}$48.7$^{s}$, $\delta$J2000 = 18$\degree$10\arcmin09.8\arcsec. The spectrum presented in Figure \ref{mom}a was extracted from the data cube using the \textit{CLEAN} mask and shows a clear double peaked structure, also consistent with emission arising from Keplerian rotation. 

Figure \ref{mom}b displays the dust continuum. We used the same restoring beam as applied to the H$_2$CO emission to enable comparison of dust (representative of the overall disk density gradient) and line emission profiles. The known inner disk dust cavity of 19 AU \citep{Espaillat_2010, Andrews_2011} is unresolved. An integrated moment-0 map of the H$_2$CO emission is shown in Fig. \ref{mom}c, displaying a large H$_2$CO emission disk without any apparent hole or gap. The emission extends more than a full beam-width beyond the expected CO snowline location from chemical models (see Section \ref{model}). The moment map was created using the \textit{immoments} command in CASA, summing over all channels observed to have H$_2$CO emission (4-8 km s$^{-1}$).

To analyze the H$_2$CO distribution, we created radial profiles (Fig. \ref{mom}d) by deprojecting the observed emission along a principal axis of -24.3$\degree$ with an inclination of 34$\degree$ \citep{Teague_2015} and azimuthally averaging about the continuum peak. These profiles do not directly translate to column density due to optical depth and excitation effects, but provide a useful visualization.  The continuum profile is shown in black and the observed H$_2$CO emission in blue. The H$_2$CO emission is centrally peaked, but exhibits a broad radial distribution compared to the disk density profile, with a half-light radius of $\sim$270.

\subsection{Chemical modeling\label{model}}

\begin{figure*}[ht!]
\centering
\includegraphics{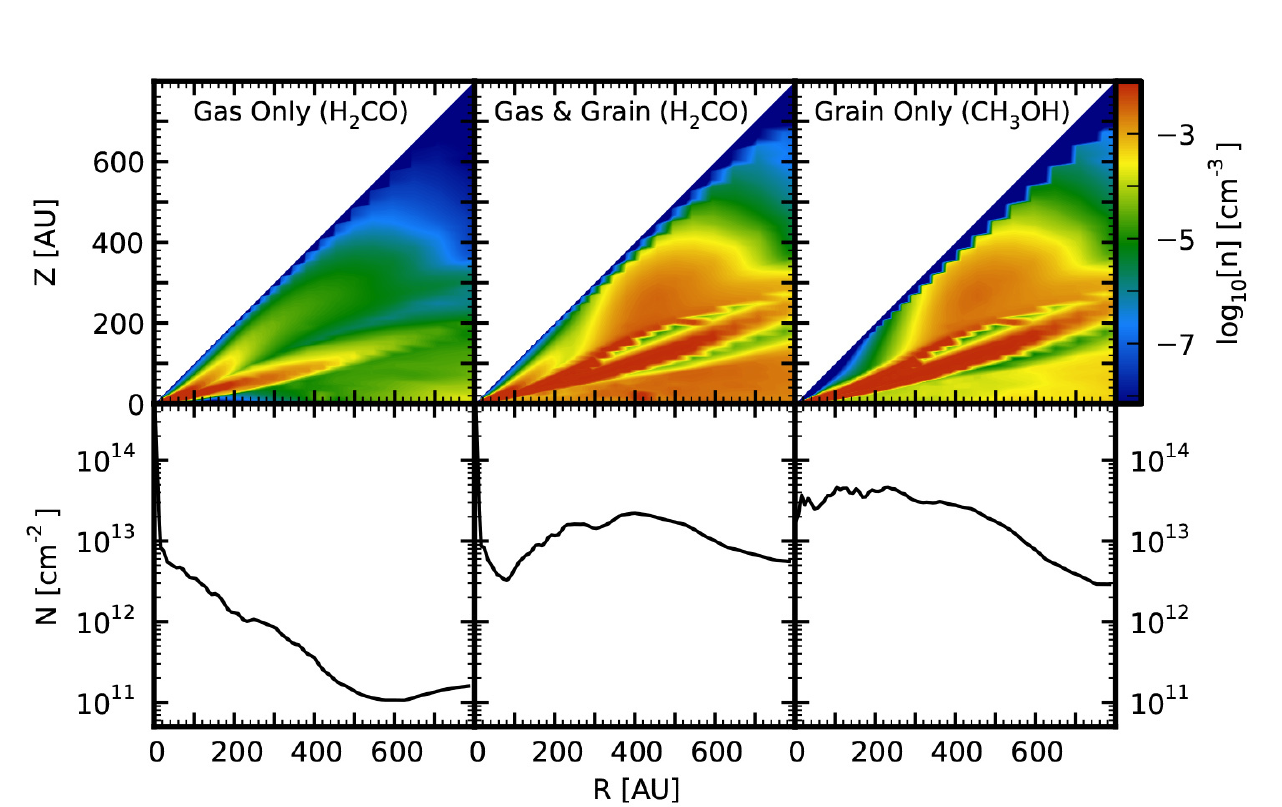}
\caption{{\small Model predictions for the H$_2$CO gas-phase density structures (top) and column densities (bottom). Models are shown with (middle column) and without (left) grain-surface chemistry. We also show  predictions for CH$_3$OH (right column), an almost pure grain-surface chemistry product. \textit{Top:} Calculated molecular abundances at various (R,Z) positions within the disk. \textit{Bottom:} Integrated column density as a function of radius.}\label{abund}}
\end{figure*}

To aid interpretation of the observations we ran a small set of disk chemistry models. We use the physical disk model of DM Tau derived in \cite{Andrews_2011} and the time-dependent chemical model presented in \cite{Cleeves_2014}, assuming 1 Myr of chemical evolution. The initial chemical abundances of the model are listed in Table 1. The high energy stellar FUV and X-ray radiation field within the disk is calculated using the Monte Carlo code and cross sections from \cite{Bethell_2011a}, see \cite{Cleeves_2013, Cleeves_2014} for details.  The stellar UV spectrum is taken to be the observed TW Hya FUV spectrum from \cite{Herczeg_2002, Herczeg_2004}, scaled to the brighter DM Tau FUV luminosity by a factor of 2.16 \citep{Yang_2011}. Finally, DM Tau does not have a measured X-ray luminosity, so the X-ray upper limit is assumed from \cite{Damiani_1995} of $L_{\rm XR} \le 4.6\times10^{29}$ erg~s$^{-1}$ and the quiescent spectral template presented by \cite{Cleeves_2013}. The standard cosmic ray ionization rate is assumed \citep[e.g. $\zeta_{\rm CR} \sim (3-7)\times10^{-17}$ s$^{-1}$,][]{Black_1990}, but since this is not directly constrained for DM Tau, we also run the models with a reduced rate and compare the results.

\begin{table}[ht!]
\begin{center}
\small
\begin{threeparttable}[b]
\caption{Chemical model initial abundances}
\begin{tabular}{+c^c^c^c}
\toprule
	Species & Abundance$^a$ &  Species  & Abundance$^a$ \\
	\otoprule
	H$_2$ & $5.00\times10^{-1}$ 		& H$_2$O(gr) & $2.50\times10^{-4}$ \\
	He & $1.40\times10^{-1}$ 			& N & $2.25\times10^{-5}$ \\
	CN & $6.00\times10^{-8}$ 		& H$_{3}^{+}$ & $1.00\times10^{-8}$ \\
	CS & $4.00\times10^{-9}$ 		& SO & $5.00\times10^{-9}$ \\
	Si$^+$ & $1.00\times10^{-11}$ 		& S$^+$ & $1.00\times10^{-11}$ \\
	Mg$^+$ & $1.00\times10^{-11}$ 	& Fe$^+$ & $1.00\times10^{-11}$ \\
	C$^+$ & $1.00\times10^{-9}$ 		& CO & $1.00\times10^{-4}$ \\
	N$_2$ & $1.00\times10^{-6}$ 		& C & $7.00\times10^{-7}$ \\
	NH$_3$ & $8.00\times10^{-8}$ 	& HCN & $2.00\times10^{-8}$ \\
	HCO$^+$ & $9.00\times10^{-9}$ 	& C$_2$H & $8.00\times10^{-9}$ \\
    \bottomrule
\end{tabular}
\begin{tablenotes}
\item[a] Abundances are relative to the proton density $n_p = 2 n_{H_{2}}$.
\end{tablenotes}
\end{threeparttable}
\end{center}
\end{table}

The central grain-surface reactions in the network include the formation of H$_2$, H$_2$O, CH$_4$, NH$_3$, H$_2$CO, CH$_3$OH and CO$_2$. For the grain-surface reactions, we assume that $10\%$ of the time the grain-surface reactions result in desorption of the product \citep[e.g.,][]{Garrod_2013}, and as this rate has not been directly measured, we also compare to the results of a $1\%$ efficiency. In the gas-phase, the primary formation route for H$_2$CO is O + CH$_3$, and thus only proceeds in the upper layers of the disk where oxygen atoms are present in the gas. These oxygen atoms are mainly supplied by the dissociation of gas-phase CO via X-ray generated He$^+$. At most locations in the disk, this particular H$_2$CO-producing reaction is more efficient than any other gas-phase formation pathways for H$_2$CO by an order of magnitude or more. The H$_2$CO and CH$_3$OH binding energies assumed in the present models are 2050 K \citep{Garrod_2006} and 4930 K \citep{Brown_2007}, respectively. Consequently, at all heights in the disk outside of $\sim100$ AU, both of these species are only non-thermally desorbed.  In the surface layers, where $z/r > 0.25$, direct UV photodesorption of H$_2$CO and CH$_3$OH ices are the primary non-thermal desorption mechanisms responsible for the gas phase abundances.  Below this layer, reactive desorption of both H$_2$CO and CH$_3$OH are the dominant return processes.  The full network includes only the standard isotopes of C, N, O, and H (e.g., no deuterated isotopologues) and contains a total of $\sim6000$ reactions and $\sim600$ species.

For comparison to the observations, we consider three different H$_2$CO formation scenarios: (1) gas-phase only; (2) both gas-phase and grain-surface; and (3) grain-surface reactions only. Calculations for the first two of these scenarios were performed by turning on and off the grain-surface reactions for H$_2$CO in the \citet{Cleeves_2013} network. It is not possible to turn off H$_2$CO gas-phase formation and destruction pathways in the model while leaving reactions on the grains because of the close coupling between H$_2$CO gas-phase chemistry and the production of many other chemical species. For this reason, CH$_3$OH \citep[which is dominated by grain-surface formation, e.g.][]{Hidaka_2004}) was used to simulate a pure grain-surface production scenario. 
Figure \ref{abund} shows the resultant radial and vertical abundance profiles of each model, as well as the integrated column density profiles. The gas-phase only model yields a very centrally peaked column density profile; the column density drops by more than an order of magnitude inside 100 AU. In contrast, the grain-surface only model column density is almost flat out to 400 AU followed by a slow decline. In the gas and grain-surface formation model, gas-phase formation is visible in the inner disk, with a sharp central peak, but outside the CO snowline at $\sim$80 AU there is a rise in column density due to grain-surface formation. As expected, gas-phase formation regulates the H$_2$CO abundance in the hot and dense inner disk, while the inclusion of grain-surface reactions increases abundances outside the CO snow line, where CO has condensed out onto grains and is available for H$_2$CO formation.

In addition to radial differentiation, the model results show that there is vertical structure in the H$_2$CO abundances (top panels of Fig. \ref{abund}). Grain-surface formation of H$_2$CO mainly enriches the disk in H$_2$CO gas in a banded layer where z/r is $\sim$0.25. This can be explained as a chemical balancing act, where the grain-surface formation and desorption of H$_2$CO is most efficient at the CO ``snow-surface" in the presence of X-rays (or other ionizing radiation). The grain-surface reactions to form H$_2$CO and CH$_3$OH in \cite{Cleeves_2013} rely on atomic hydrogen additions, where H formation is almost entirely driven by the ionization of H$_2$ by X-rays from the cool central star. Above this snow-surface layer CO is not frozen out in high concentrations, and below this layer the disk becomes optically thick to X-rays, making the grain-surface chemistry inefficient.

To explore the robustness of these results to the chemical model inputs that have not been directly measured, we ran two additional models with a reduced cosmic ray ionization rate of $\zeta_{\rm CR} \sim (1)\times10^{-19}$ s$^{-1}$ and a reduced reactive desorption efficiency of 1\%, respectively. A reduced cosmic ray ionization rate, as might be present in a disk with a "T-Tauriosphere'' generated by stellar winds \citep{Cleeves_2013}, reduces the total amount of grain surface H$_{2}$CO production, particularly in the mid-plane where other forms of radiation are not able to penetrate. Reactive desorption similarly affects mainly midplane abundances, where it is the main ice desorption pathway; at higher disk layers UV photodesorption is a more important desorption mechanism. 
Compared to the standard chemical model above, the models with reduced cosmic ray ionization or reactive desorption result in a substantial (order of magnitude) reduction in H$_2$CO gas in the disk midplane. These abundance reductions in the midplane have no significant effect on the emission profile, however, as most of the emission comes from the relatively unaffected z/r $\sim$0.25 surface rather than the midplane. The predicted radial H$_2$CO emission profiles for the three H$_2$CO formation scenarios displayed in Fig \ref{abund} thus seem robust with respect to the assumed cosmic ray ionization rate and reactive desorption efficiency.

\subsection{Simulated observations of chemical models}

\begin{figure}[ht!]
\centering
\includegraphics[scale=0.9]{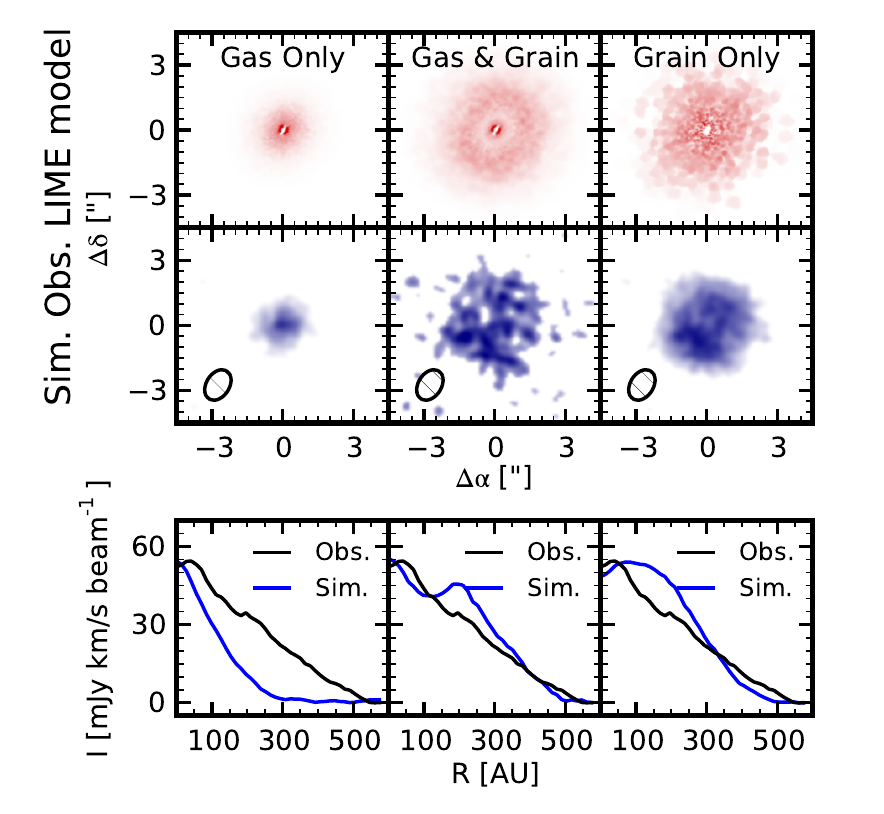}
\caption{{\small Simulations of H$_2$CO emission toward DM Tau, using the molecular abundance structures from chemical models shown in Fig. \ref{abund}. \textit{Top:} Non-LTE radiative transfer simulation of the 5$_{15}$ - 4$_{14}$ line of H$_2$CO using LIME. \textit{Middle:} Simulated ALMA observations scaled to the peak observed flux, with the synthesized beam shown in the lower left. \textit{Bottom:} Radial emission profiles, where simulated observations have been deprojected and azimuthally averaged similarly to the data.
The simulated emission is shown in blue and observed emission shown in black.}\label{sim}}
\end{figure}

From the modeled molecular abundance structures, we calculated emission profiles of H$_2$CO 5$_{15}$ - 4$_{14}$ using the non-LTE radiative transfer code LIME \citep{Brinch_2010}. A stellar mass of 0.7 M$_{sun}$ (to determine line broadening), inclination of 34$\degree$, and distance of 145 pc were assumed for the radiative transfer \citep{Teague_2015}.  The LIME simulations are shown in the top panels of Fig. \ref{sim}.  ALMA observations were then simulated in CASA using \textit{simobserve}, assuming the antenna configuration and integration time of the original observations (see Section \ref{observe}). The original model output overpredicted the line fluxes by factors of 2--10. To enable comparisons of observed and simulated radial profiles, the model abundances were uniformly scaled such that the peak integrated flux of the model and the observations match (middle and bottom panels of Fig. \ref{sim}). The model over-predictions are not surprising, since H$_2$CO and CH$_3$OH are at the maximum extent of the complexity considered in the reaction network, and thus there are limited reaction pathways away from these species. Additionally, as the destruction products of H$_2$CO and CH$_3$OH have a limited chemistry and are not able to form more complex species, H$_2$CO or CH$_3$OH re-creation events are likely. Observations have also shown that CO abundances in disks may be depleted and time evolving, leading to possible over-prediction of molecules forming from CO in the model \citep{Favre_2013}.

The three models exhibit very different emission profiles. The gas-phase only model yields compact centrally peaked emission. Both models including grain-surface reactions have substantial H$_2$CO emission out to several hundred AU. When comparing observed and simulated radial profiles, it is clear that the gas-phase model severely underpredicts emission in the outer disk. By contrast, grain-surface formation alone overpredicts emission outside 100 AU. The gas and grain-surface chemistry model is most similar to the observations, but still overpredicts the relative amount of H$_2$CO emission between 150 and 300 AU compared to the emission from the disk center.

\section{Discussion}

The presented ALMA observations show that the H$_2$CO emission around DM Tau is fairly extended, consistent in radial extent with the observations of \cite{Henning_2008}. In contrast to their detected ring shape, we find that H$_2$CO is centrally peaked. The detection of H$_2$CO emission from the inner disk in this data set and its absence in \cite{Henning_2008} can be ascribed both to a combination of higher sensitivity and better uv coverage in the ALMA data, enabling us to recover weaker emission, and the fact that the two transitions have different excitation properties. In particular, the H$_2$CO transition analyzed in this work (5$_{15}$ - 4$_{14}$) has a significantly higher excitation energy than the 3$_{13}$ - 2$_{12}$ transition detected by \cite{Henning_2008} (E$_u$ = 62.45 vs 32.06 K)\footnote{Taken from www.splatalogue.net \citep{Remijan_2007_splat}. Lab data used in fit are from \cite{Brunken_2003}}, and thus the previously noted H$_2$CO ring may be partially due to excitation effects. Supporting this hypothesis, LIME simulations of our gas and grain model for the 3$_{13}$ - 2$_{12}$ transition show a substantially reduced inner disk component (not shown) compared to the 5$_{15}$ - 4$_{14}$ transition.

The emission profile we observe in the outer disk of DM Tau (i.e. outside the CO snow line) can be explained well by grain-surface formation and desorption of H$_2$CO. However an explanation that accounts for the centrally peaked emission line profile also demands a contribution from gas-phase reactions. The proposed scenario where both gas and grain-surface chemistry contribute significantly to the disk H$_2$CO emission is confirmed by chemical modeling. The model results also demonstrate that the relative gas and grain-surface chemistry contributions to the observed H$_2$CO gas are highly dependent on disk radius. 

\section{Conclusions and future directions}

Based on ALMA observations and detailed chemical modeling, we have shown that there are both gas-phase and grain-surface contributions to the observed H$_2$CO gas in DM Tau. This is the first evidence of gas-phase formation of H$_2$CO toward a protoplanetary disk, and implies that H$_2$CO cannot be used as a simple tracer of grain-surface chemistry without constraining the relative contributions of each formation route. 

The exact morphology of the H$_2$CO emission will depend sensitively on the ratio of these contributions, but will be difficult to recover from observations due to excitation degeneracies (as demonstrated in the \cite{Henning_2008} observations) and the currently unknown vertical abundance structure. Additional ALMA observations of multiple H$_2$CO transitions could be used to provide constraints on excitation conditions, and therefore the vertical H$_2$CO structure in a disk. Such observations would also provide important constraints on ice desorption from grain-surfaces and thus how well ice and gas abundances are coupled in disks \citep[e.g.][]{Oberg_2007, Fayolle_2011}.

Finally, we have conclusively shown that some fraction of the H$_2$CO around DM Tau must originate from hydrogenation of CO on grain-surfaces. We therefore expect the next hydrogenation product of H$_2$CO, methanol (CH$_3$OH), to be present as well \citep{Cuppen_2009, Walsh_2014}. Although it has yet to be detected toward a protoplanetary disk, searching for both CH$_3$OH and the complex products of its photochemistry \citep[e.g][]{Garrod_2013, Elsila_2007, Thi_2004} clearly warrants further exploration as the capabilities of ALMA continue to improve.  

\acknowledgments

We thank the anonymous referee for useful comments. We thank David Wilner and Charlie Qi for productive and insightful conversations.  RAL gratefully acknowledges funding from an National Science Foundation Graduate Research Fellowship.  LIC acknowledges support by NSF grant AST-1008800 and the Rackham Predoctoral Fellowship. KIO also acknowledges funding from the Simons Collaboration on the Origins of Life (SCOL), the Alfred P. Sloan Foundation, and the David and Lucile Packard Foundation. The National Radio Astronomy Observatory is a facility of the National Science Foundation operated under cooperative agreement by Associated Universities, Inc.  This paper makes use of the following ALMA data: ADS/JAO.ALMA\# 2011.0.00629.S. ALMA is a partnership of ESO (representing its member states), NSF (USA) and NINS (Japan), together with NRC (Canada) and NSC and ASIAA (Taiwan), in cooperation with the Republic of Chile. The Joint ALMA Observatory is operated by ESO, AUI/NRAO and NAOJ.

\end{document}